\documentclass[APL,reprint,superscriptaddress,showpacs]{revtex4-1}

\usepackage{amsmath}    
\usepackage{graphicx}   
\usepackage{color}      
\usepackage{subfigure}  
\usepackage{hyperref}   
\usepackage{todonotes}
\usepackage[latin1]{inputenc}

\begin{document}

\title{Site-specific probing of charge transfer dynamics in organic photovoltaics}
\author{Tiberiu Arion}
\affiliation{Center for Free-Electron Laser Science / DESY, Notkestra\ss e 85, D-22607 Hamburg, Germany}

\author{Stefan Neppl}
\affiliation{Chemical Sciences Division, Lawrence Berkeley National Laboratory, Berkeley, California 94720, USA}

\author{Friedrich Roth}
\affiliation{Center for Free-Electron Laser Science / DESY, Notkestra\ss e 85, D-22607 Hamburg, Germany}

\author{Andrey Shavorskiy}
\affiliation{Chemical Sciences Division, Lawrence Berkeley National Laboratory, Berkeley, California 94720, USA}

\author{Hendrik Bluhm}
\affiliation{Chemical Sciences Division, Lawrence Berkeley National Laboratory, Berkeley, California 94720, USA}

\author{Zahid Hussain}
\affiliation{Advanced Light Source, Lawrence Berkeley National Laboratory, Berkeley, California 94720, USA}

\author{Oliver Gessner}
\affiliation{Chemical Sciences Division, Lawrence Berkeley National Laboratory, Berkeley, California 94720, USA}

\author{Wolfgang Eberhardt}
\affiliation{Center for Free-Electron Laser Science / DESY, Notkestra\ss e 85, D-22607 Hamburg, Germany}
\affiliation{Advanced Light Source, Lawrence Berkeley National Laboratory, Berkeley, California 94720, USA}
\affiliation{Institute of Optics and Atomic Physics, TU Berlin, Stra\ss e des 17. Juni 135, D-10623 Berlin, Germany}
\date{\today}


\begin{abstract}
We report the site-specific probing of charge-transfer dynamics in a prototype system for organic photovoltaics (OPV) by picosecond time-resolved X-ray photoelectron spectroscopy. A layered system consisting of approximately two monolayers of C$_{60}$ deposited on top of a thin film of Copper-Phthalocyanine (CuPC) is excited by an optical pump pulse and the induced electronic dynamics are probed with 590\,eV X-ray pulses. Charge transfer from the electron donor (CuPC) to the acceptor (C$_{60}$) and subsequent charge carrier dynamics are monitored by recording the time-dependent C\,1$s$ core level photoemission spectrum of the system. The arrival of electrons in the C$_{60}$ layer is readily observed as a completely reversible, transient shift of the C$_{60}$ associated C\,1$s$ core level, while the C\,1$s$ level of the CuPC remains unchanged. The capability to probe charge transfer and recombination dynamics in OPV assemblies directly in the time domain and from the perspective of well-defined domains is expected to open additional pathways to better understand and optimize the performance of this emerging technology.
\end{abstract}

\maketitle

Research on materials for organic photovoltaic (OPV) applications is driven mostly by two figures of merit: the light-to-current conversion efficiency and the cost of production, which would make the technology suitable for widespread implementation. So far, OPVs meet the second criterion, as they can easily be prepared by deposition from solutions, or by low-cost printing techniques, rather than using more demanding vacuum deposition methods. However, the relatively low conversion efficiency of OPVs remains a significant challenge. It is generally accepted that recombination of the photo-induced electron-hole pair or excitonic state in the photoreceptive material is one of the major loss mechanisms. Consequently, many research efforts focus on the identification of methods to quench the recombination process. Empirically, the admixture of C$_{60}$ into organic photoreceptor films has led to significant progress in this respect \cite{Morita1992,Sariciftci1992,Schlebusch1996,Schlebusch1999,Morenzin1999,Kessler1998,Wilke2010,Schlebusch_thesis}. Based on observations of intrinsic fluorescence quenching \cite{Morita1992,Sariciftci1992}  as well as from photoemission studies of the band alignment in thin film systems \cite{Gunnarsson1995}, it has been concluded that after the initial electron-hole production in the chromophore, the electron is captured by the C$_{60}$ molecule, thus spatially separating the electron from the hole in the chromophore molecule. This spatial separation prevents direct electron-hole recombination. Studies by photoelectron spectroscopy (PES) have shown that the charge transfer process is energetically possible, both in layered systems \cite{Schlebusch1999, Morenzin1999} as well as in heterogeneous mixtures \cite{Roth2014}. Here, we concentrate on the material combination of the organic photoreceptor copper phthalocyanine (CuPC), which is a ubiquitous component of photochemical and light harvesting applications \cite{Morita1992,Sariciftci1992,Schlebusch1996}, with C$_{60}$ molecules as electron acceptors.

A recent report on CuPC-C$_{60}$ interfaces indicates that, when CuPC is deposited atop of a C$_{60}$ film, the sample surface tends to charge up, unlike the case of a C$_{60}$ film grown on top of a CuPC covered substrate\cite{Wilke2010}. Detailed overviews of some of the challenges encountered in applying photoelectron spectroscopy to organic-based photovoltaics (e.g. radiation damage, charging effects, errors in determining the binding energy of the hole-transporting species) and ways to tackle these effects can be found in \cite{Opitz2013} and \cite{Fahlman2013}. The current state of research on organic heterojunctions and their implications for the development of organic solar cells has recently been addressed by several authors based on experimental results \cite{Kushto2010,Akaike2010,Koch2012,Li2014} and theoretical approaches \cite{Castet2014,Beltran2014,Sai2012}. In particular, Dutton and Robey \cite{Dutton2012} and Jailaubekov et al. \cite{Jailaubekov2013} have investigated the dynamics of hot charge-transfer excitons in C$_{60}$/phthalocyanine heterojunctions by time-resolved two-photon-photoemission (TR-2PPE). However, by contrast to valence band PES, which is not necessarily site-specific with respect to the spatial location of the excited electronic states, XPS is, making such an approach a valuable complement to existing experiments.

\par

 While the model of improved photovoltaic performance by charge separation at the CuPC-C$_{60}$ interface is quite plausible, so far no direct observation of the electron transfer to the C$_{60}$ domains has been reported. Once a clear signature of the electron's presence at the C$_{60}$ can be identified, monitoring the timescale and efficiency of the charge transfer process offers an excellent possibility to boost the performance of this and related OPV systems. The probability and timescale of the charge transfer step is expected to depend on the morphology of the system in addition to the energetic alignment. Here, we demonstrate that time-resolved XPS can be used to directly observe the transfer of electrons to the C$_{60}$ acceptor molecules, and to access the relevant timescales of the charge transfer and relaxation processes.

\par

The time-resolved XPS experiment was performed at Beamline 11.0.2 of the Advanced Light Source (ALS), using the HPPES (High Pressure Photoemission Spectroscopy) endstation \cite{Bluhm2006,Neppl2014}. The ALS was operated in two-bunch filling mode with a bunch-to-bunch spacing of 328\,ns and a pulse length of $\sim$\,70\,ps \cite{Glover2001}. The pump laser system (Time Bandwidth Products DUETTO) provided 10\,ps pulses at a wavelength of 532\,nm. It was operated at a repetition rate of 126.9\,kHz and synchronized to the ALS X-ray pulse train \cite{Neppl2014}. The samples were prepared by in situ sequential evaporation on a pre-cleaned Si(100) wafer (n-doped) and consisted of approximately two monolayers of C$_{60}$ deposited on top of a thin CuPC film. The C$_{60}$ layer thickness was estimated by the relative intensities of the CuPC and C$_{60}$ XPS signals, similar to earlier reports \cite{Schlebusch1999}. The laser spot size was 450\,$\mu$m\,x\,450\,$\mu$m with a synchrotron beam spot size of 70\,$\mu$m\,x\,70\,$\mu$m, ensuring spatial overlap and homogeneous excitation conditions. The temporal overlap was calibrated using the ultrafast surface photovoltage (SPV) response of the Si(100) substrate \cite{Neppl2014}. The sample was scanned continuously during the experiment in order to avoid any impact of sample damage on the reported results.

\begin{figure}[t]
\includegraphics[width=0.8\linewidth]{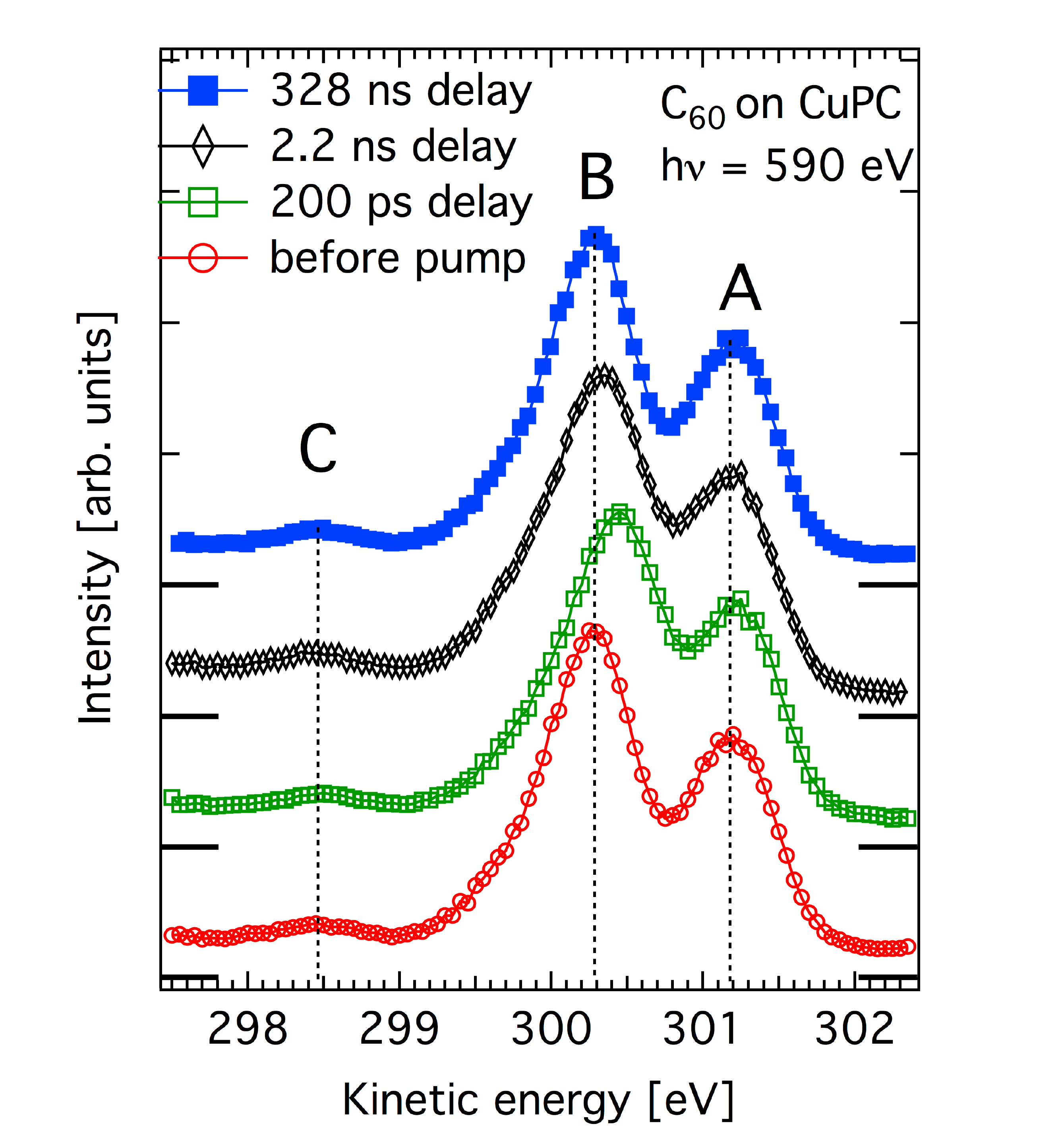}
\caption{Time-resolved XPS spectra of a bi-layer system consisting of $\sim$\,2 monolayers of C$_{60}$ deposited on top of a thin film of CuPC ($h\nu$\,=\,590\,eV). The red curve (open circles) shows the XPS spectrum before the pump pulse. The green (open squares), black (open diamonds), and blue curves (full squares) show the XPS spectra recorded 200\,ps, 2.2\,ns, and 328\,ns, respectively, after the pump pulse arrival. The vertical dotted black lines have been added to guide the eye. They indicate the locations of the maxima of the three spectral features for the unexcited system.}
\label{f1}
\end{figure}

\begin{figure}[t]
\includegraphics[width=1.01\linewidth]{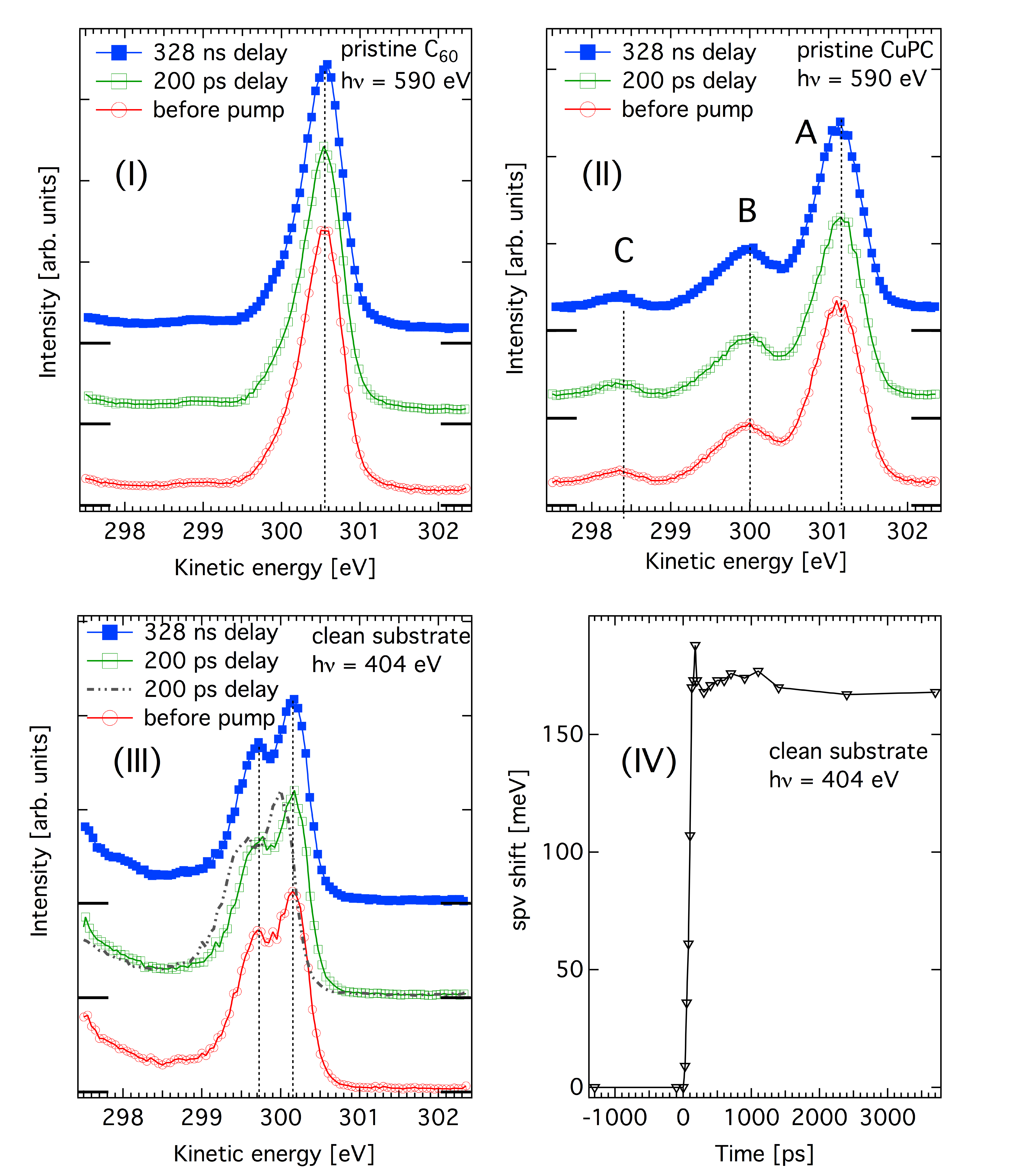}
\caption{Time-resolved XPS spectra of thin films of pristine C$_{60}$ (I) and pristine CuPC (II) deposited on a pre-cleaned Si wafer (n-doped), and of a clean substrate (Si wafer) (III). The red curve (open circles) shows the XPS spectrum before the pump pulse arrival, while the green (open squares) and the blue curves (full squares) show the probe spectra recorded 200\,ps and 328\,ns after optical excitation, respectively. (IV) Time evolution of the SPV of the clean Si wafer. The grey (dotted line) curve in Fig.\,\ref{f2} (III) shows the uncorrected XPS spectrum of the substrate recorded 200 ps after the laser pulse has reached the sample.}
\label{f2}
\end{figure}

Figure\,\ref{f1} shows the time resolved XPS spectra of the bi-layer system. The red curve and open circles show the XPS spectrum recorded when the X-ray pulse precedes the 532\,nm (1\,mJ/cm$^2$) pump pulse. The green (open squares), black (open diamonds), and blue curves (full squares) show the XPS probe spectra recorded 200\,ps, 2.2\,ns, and 328\,ns, respectively, after the pump pulse arrival. We note that the optical excitation creates electron hole pairs in the CuPC but not in the C$_{60}$, as demonstrated in previous UV-Vis absorption experiments of the individual compounds \cite{Laurs1987}. The vertical dotted (black) lines have been added to guide the eye, indicating the locations of the maxima of the three spectral features for the unperturbed system. The spectra recorded after the pump pulse arrival have been corrected for the surface photovoltage shift (SPV) of the Si substrate, which also shifts the CuPC emission of the deposited film, as discussed in detail below. The curves are offset vertically for improved clarity as indicated by the longer tickmarks. We observe significantly different dynamics for different spectral features. While features A and C are unaffected by the optical excitation, feature B exhibits a distinct time dependent, reversible shift at 200 ps and 2.2\,ns delays, and has fully recovered to its original position after 328\,ns. Based on earlier XPS studies \cite{Kessler1998} and the spectra of the isolated systems shown in Fig.\,\ref{f2} (I), feature B is largely assigned to the C\,1$s$  XPS from C$_{60}$, superimposed upon a weaker emission feature from CuPC. Features A and C are spectrally pure components of the C\,1$s$  XPS from CuPC (Fig.\,\ref{f2} (II)). The time-dependent shift of B upon laser excitation is, therefore, a signature of charge transfer from the chromophore CuPC to the electron acceptor C$_{60}$. A more detailed picosecond time-resolved investigation of the core photoemission dynamics is in planning, and will be the focus of a future study.

\begin{figure}[t]
\includegraphics[width=1.01\linewidth]{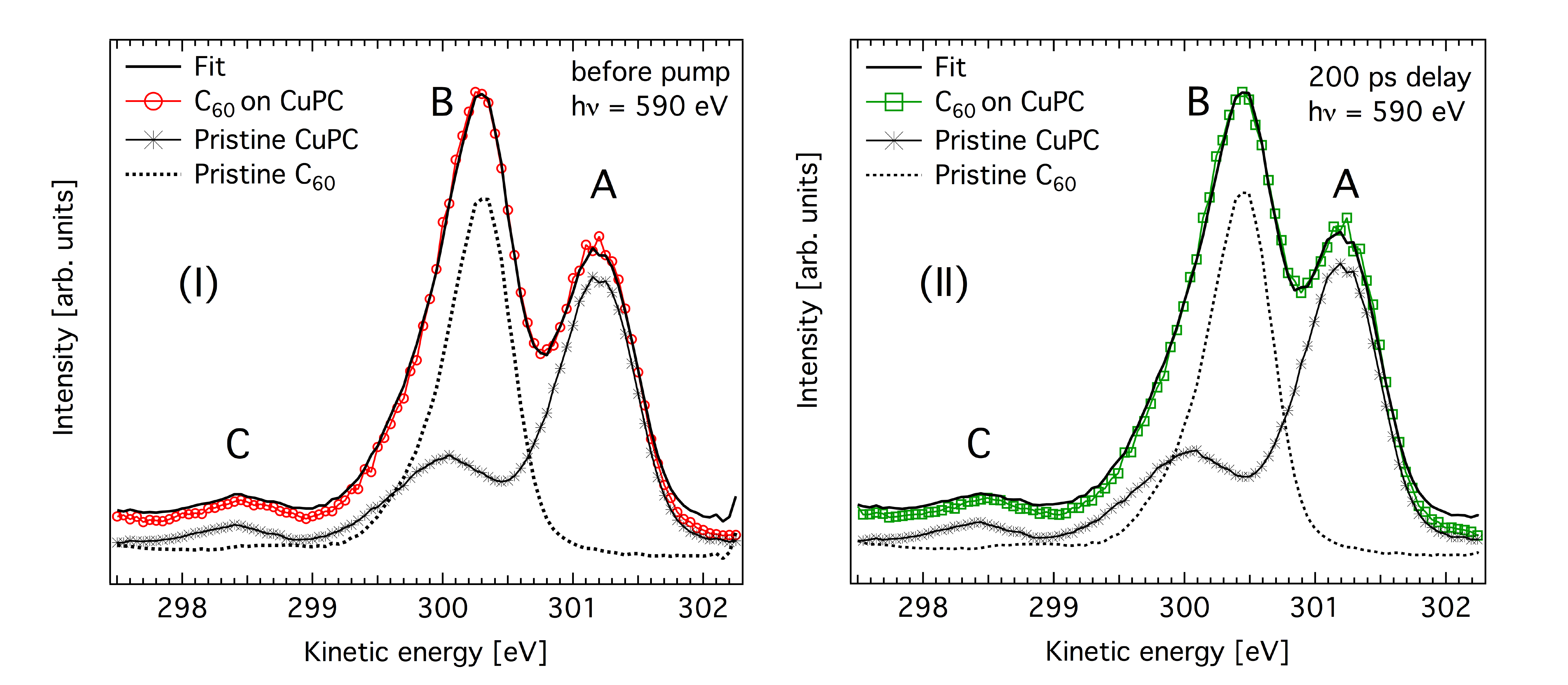}
\caption{XPS spectra ($h\nu$\,=\,590\,eV) of the bi-layer system of C$_{60}$ on CuPC (open circles, red curve and open squares, green curve) as recorded before the pump pulse arrival (I) and 200\,ps after the optical excitation (II), respectively. The spectrum of the excited bi-layer has been corrected for SPV (about 160\,meV) and is displayed on top of the recorded spectra of pristine CuPC (stars, full black curve) and C$_{60}$ (dashed, black line). The black full line indicates the result of a fit that describes the experimental spectra of the bi-layer system by a linear combination of the single layer spectra.}
\label{f3}
\end{figure}

In order to verify the spectral assignments we have conducted several test experiments on pure CuPC films, pure C$_{60}$ films, and on the bare Si (100) substrate. Figures\,\ref{f2} (I) and (II) show the time-resolved C\,1$s$ spectra of thin films of pristine C$_{60}$ and pristine CuPC, respectively, deposited on pre-cleaned, n-doped Si wafers. Time-resolved Si\,2$p$ spectra of the clean Si substrate are shown in Fig.\,\ref{f2} (III). The red curves in Figs.\,\ref{f2} (I),(II),(III) (open circles) show the XPS spectra before the pump pulse arrival, while the green (open squares) and blue curves (full squares) show the probe spectra recorded 200 ps and 328\,ns, respectively, after the optical excitation. The peaks in Fig.\,\ref{f2} (II) are labeled following the notation in \cite{Kessler1998}. The photoelectron spectra recorded after the pump pulse has arrived at the sample have been corrected for the transient surface photovoltage (SPV) of the Si substrate  \cite{Long1990}. The baselines of the spectra taken after the pump pulse are shifted vertically as indicated by the longer tickmarks. No changes in the spectral shapes or peak positions beyond the Si SPV response are observed for either of the pristine samples or the clean substrate upon pumping with 532\,nm light. For improved clarity, the uncorrected XPS spectrum of the substrate recorded 200\,ps after the laser pulse has reached the sample is included as grey, dashed curve in Fig.\,\ref{f2} (III).

\par

Figures\,\ref{f2} (IV) shows the time evolution of the SPV in the clean Si substrate. The SPV is determined by using the Si\,2$p$ XPS spectra shown in Fig.\,\ref{f2} (III). The perfect alignment of the corrected spectra corroborates the reproducibility of the measurements and the reliability of the corrections. Upon pumping with the optical laser, a transient SPV of about 160\,mV is observed within the response time of the apparatus. Over the timescale of the experiment (3.5\,ns) the system relaxes only to a small extent. However, within this time interval, the time dependent shift of peak B in Fig.\,\ref{f1} has almost completely recovered to its initial position, demonstrating that the dynamics of peak B are driven by a different mechanism than the substrate SPV dynamics. Traces recorded for longer pump-probe time delays (not shown) reveal that the surface potential has fully recovered after 3.6\,$\mu$s, i.\,e. well before the next optical pump pulse arrives.

\par

It is important to emphasize that upon laser irradiation the spectra of the pure films show no other change than the SPV shift of the Si substrate. Apart from this rigid shift, the line shapes and peak positions are identical.

\par

In order to quantify the observed trends in Fig.\,\ref{f1}, a component analysis of the time resolved XPS spectra is performed as illustrated in Figs.\,\ref{f3} (I) and (II). They show the XPS spectra of the bi-layer system of C$_{60}$ on CuPC (open circles, red curve and open squares, green curve) recorded (I) before and (II) 200\,ps after the optical excitation, respectively. The color convention is the same as in Fig.\,\ref{f1}. The measured spectra can be well described by linear combinations of the spectra of pristine CuPC (stars, black curve) and C$_{60}$ (dashed black line) spectra, which have been shifted and scaled in a least-squares fit procedure to give the best representation of the time-dependent XPS spectra of the combined system (black full curve) \cite{Schlebusch1996,Schlebusch_thesis,Roth2014}. Note that the composite curves describe the experimental results very well despite the simplicity of the four-parameter fits (2 energy shifts and 2 amplitudes).

The analysis shows that the C\,1$s$ XPS of CuPC does not change upon optical excitation, whereas the C\,1$s$ spectrum of C$_{60}$ on top of CuPC exhibits a reversible time dependent shift of 144\,meV to lower binding energies. This shift is compatible with the presence of an additional negative screening charge on C$_{60}$, which has been injected from the CuPC.

In other words, upon optical laser pumping, the CuPC molecules are excited and electrons in the HOMO of the CuPC molecules are promoted into an excitonic state, while the C$_{60}$ molecules are unaffected. However, due to the high electron affinity of the fullerenes, the electrons rapidly move towards the C$_{60}$ layer, whereby the exciton dissociates to inject an electron into C$_{60}$ and a hole remains in the CuPC phase. Thus, a large number of electrons have arrived in the C$_{60}$ layer 200\,ps after the optical excitation, and C$_{60}$ molecules immediately attract an electron from their vicinity upon core ionization. This does not mean that an additional electron is attached to every single C$_{60}$ molecule, but rather that upon core ionization, efficient core-hole screening is enabled by charge transfer within the C$_{60}$ or by hopping from neighboring molecules. With increasing time delay occurs a reduction of the free charge carrier concentration, such that, on average, screening of the fullerene core holes is reduced, i.\,e., the spectral signature of the reduced charge carrier density is an intermediate shift value of the XPS feature B. We, therefore, interpret the intermediate shifts as a superposition of screened and unscreened emission peaks, which are not resolved in our data, and which exhibit time-dependent relative intensities. Once the electron-hole recombination process is complete, the system has returned to its original state as observed before the pump pulse arrival.

Note that a similar shift as observed here for feature B has previously been detected in alkali-doped fullerenes \cite{Poirier1995}. In terms of XPS terminology it is predominantly a final state shift. The additional electron density on the C$_{60}$ is highly mobile and readily screens any core holes, irrespective of their location within the C$_{60}$ film. 

\par

We emphasize that the observed effect does not require one electron to be present on each C$_{60}$ molecule. The bandwidth of the LUMO of the C$_{60}$ bilayer is about 500\,meV \cite{Haddon1992}, whereas the lifetime broadening of a C\,1$s$ core hole in C$_{60}$ is about 70\,meV. Accordingly, the charge transfer time within the C$_{60}$ film is about an order of magnitude faster than the core hole lifetime. This is the reason why electron screening of core holes created in the C$_{60}$ film is (initially) observed with high probability, and this probability decreases as the total amount of free charge carriers within the C$_{60}$ film diminishes. The efficient core hole screening by the injected electrons may also indicate a very efficient electron conduction mechanism within the C$_{60}$ domains, which would be beneficial for PV applications.

\par

We would like to stress that TR-2PPE studies, e.\,g., by Dutton and Robey \cite{Dutton2012} and Jailaubekov et al. \cite{Jailaubekov2013} led to the identification of several intramolecular and interfacial excitonic states. A dissociated excitonic state, however, and the timescales for charge transfer to the C$_{60}$ layer and subsequent recombination across the interface have not yet been unambiguously determined. We expect that TR-XPS studies have the potential to provide critical information in this regard.

\par

At first glance, it is surprising that the CuPC-C\,1$s$ XPS spectrum does not exhibit a similarly pronounced shift or shape change as the C$_{60}$-C\,1$s$ line upon charge transfer between the two domains. The optical excitation corresponds to a HOMO-LUMO transition with both orbitals located on the $\pi$ system of the CuPC molecule (i.e., is of $\pi-\pi^*$ character \cite{Davidson1982,Farag2007}). Therefore, the hole density after charge transfer to the C$_{60}$ domain, which is expected to be dominated by the original HOMO orbital, should be localized predominantly on the organic phthalocyanine ligands. A possible explanation for the lack of a transient CuPC-C\,1$s$ shift may lie in the combination of the particular bi-layer structure of the sample and pinning of the CuPC energy levels to the Si substrate as demonstrated by the Si SPV induced shift of the pristine CuPC XPS spectrum. Further experimental and theoretical work for various sample configurations is planned to test the viability of these assumptions.

\par

The intermolecular charge transfer process in CuPC is more than an order of magnitude slower than in the C$_{60}$ domain. This estimate is based on angle resolved photoemission data showing a lack of dispersion of the occupied electronic states in ordered CuPC films \cite{Ueno1999}. Accordingly, only the molecules containing a hole are expected to show a modified C\,1$s$ XPS signal and these modifications may be small, if detectable at all, due to the high localization of the hole densities on single molecules. Further indications that local charge dynamics in CuPC molecules may be challenging to detect are provided by XPS spectra of alkali doped CuPC films \cite{Schwieger2002}. They show that the presence of an extra charge on the CuPC molecule is much more readily detected in a shift of the Cu\,2$p$ XPS signal, while the shift is hardly observable in the C\,1$s$ signal. For a doping level of 1.7 alkali atoms/per CuPC the C\,1$s$ shift was reported to be less than 100\,meV measured relative to the valence band edge \cite{Schwieger2002}. Future studies will include the time-resolved monitoring of Cu inner-shell photolines to gain a more complete picture of charge transfer from both sides of the interface.

\par

The clear spectral signature of the charged C$_{60}$ phase opens the possibility to directly measure the charge transfer time and to monitor the temporal evolution of the charged state of the electron acceptor. We expect that these dynamic properties of the material blend depend on the electronic structure of the chromophore and on the morphology of the sample, if exciton diffusion plays the dominant role in the charge transfer process. The capability to directly test the correlation between the design of an OPV and the site-specific electron dynamics will be important in optimizing the charge generation and collection processes in a large class of solar energy and light harvesting applications.

\par

Fruitful discussions with D. S. Pemmaraju are gratefully acknowledged. We acknowledge the allocation of beam time at the beam line 11.0.2. of the ALS, as well as the support of the ALS staff during the beam time. The Advanced Light Source is supported by the Director, Office of Science, Office of Basic Energy Sciences, of the U.S. Department of Energy under Contract No. DE-AC02-05CH11231. The Molecular Environmental Sciences beamline 11.0.2 is supported by the Director, Office of Science, Office of Basic Energy Sciences, and the Division of Chemical Sciences, Geosciences, and Biosciences of the US Department of Energy at the Lawrence Berkeley National Laboratory under Contract No. DE-AC02-05CH11231. One of us (WE) would like to thank the ALS for their support and hospitality. OG was supported by the Department of Energy Office of Science Early Career Research Program. SN acknowledges support by the Alexander von Humboldt foundation.


\begin{thebibliography}{32}%
\makeatletter
\providecommand \@ifxundefined [1]{%
 \@ifx{#1\undefined}
}%
\providecommand \@ifnum [1]{%
 \ifnum #1\expandafter \@firstoftwo
 \else \expandafter \@secondoftwo
 \fi
}%
\providecommand \@ifx [1]{%
 \ifx #1\expandafter \@firstoftwo
 \else \expandafter \@secondoftwo
 \fi
}%
\providecommand \natexlab [1]{#1}%
\providecommand \enquote  [1]{``#1''}%
\providecommand \bibnamefont  [1]{#1}%
\providecommand \bibfnamefont [1]{#1}%
\providecommand \citenamefont [1]{#1}%
\providecommand \href@noop [0]{\@secondoftwo}%
\providecommand \href [0]{\begingroup \@sanitize@url \@href}%
\providecommand \@href[1]{\@@startlink{#1}\@@href}%
\providecommand \@@href[1]{\endgroup#1\@@endlink}%
\providecommand \@sanitize@url [0]{\catcode `\\12\catcode `\$12\catcode
  `\&12\catcode `\#12\catcode `\^12\catcode `\_12\catcode `\%12\relax}%
\providecommand \@@startlink[1]{}%
\providecommand \@@endlink[0]{}%
\providecommand \url  [0]{\begingroup\@sanitize@url \@url }%
\providecommand \@url [1]{\endgroup\@href {#1}{\urlprefix }}%
\providecommand \urlprefix  [0]{URL }%
\providecommand \Eprint [0]{\href }%
\providecommand \doibase [0]{http://dx.doi.org/}%
\providecommand \selectlanguage [0]{\@gobble}%
\providecommand \bibinfo  [0]{\@secondoftwo}%
\providecommand \bibfield  [0]{\@secondoftwo}%
\providecommand \translation [1]{[#1]}%
\providecommand \BibitemOpen [0]{}%
\providecommand \bibitemStop [0]{}%
\providecommand \bibitemNoStop [0]{.\EOS\space}%
\providecommand \EOS [0]{\spacefactor3000\relax}%
\providecommand \BibitemShut  [1]{\csname bibitem#1\endcsname}%
\let\auto@bib@innerbib\@empty
\bibitem [{\citenamefont {Morita}\ \emph {et~al.}(1992)\citenamefont {Morita},
  \citenamefont {Zakhidov},\ and\ \citenamefont {Yoshino}}]{Morita1992}%
  \BibitemOpen
  \bibfield  {author} {\bibinfo {author} {\bibfnamefont {S.}~\bibnamefont
  {Morita}}, \bibinfo {author} {\bibfnamefont {A.~A.}\ \bibnamefont
  {Zakhidov}}, \ and\ \bibinfo {author} {\bibfnamefont {K.}~\bibnamefont
  {Yoshino}},\ }\href {\doibase http://dx.doi.org/10.1016/0038-1098(92)90636-N}
  {\bibfield  {journal} {\bibinfo  {journal} {Solid State Commun.}\ }\textbf
  {\bibinfo {volume} {82}},\ \bibinfo {pages} {249 } (\bibinfo {year}
  {1992})}\BibitemShut {NoStop}%
\bibitem [{\citenamefont {Sariciftci}\ \emph {et~al.}(1992)\citenamefont
  {Sariciftci}, \citenamefont {Smilowitz}, \citenamefont {Heeger},\ and\
  \citenamefont {Wudl}}]{Sariciftci1992}%
  \BibitemOpen
  \bibfield  {author} {\bibinfo {author} {\bibfnamefont {N.~S.}\ \bibnamefont
  {Sariciftci}}, \bibinfo {author} {\bibfnamefont {L.}~\bibnamefont
  {Smilowitz}}, \bibinfo {author} {\bibfnamefont {A.~J.}\ \bibnamefont
  {Heeger}}, \ and\ \bibinfo {author} {\bibfnamefont {F.}~\bibnamefont
  {Wudl}},\ }\href {\doibase 10.1126/science.258.5087.1474} {\bibfield
  {journal} {\bibinfo  {journal} {Science}\ }\textbf {\bibinfo {volume}
  {258}},\ \bibinfo {pages} {1474} (\bibinfo {year} {1992})}\BibitemShut
  {NoStop}%
\bibitem [{\citenamefont {Schlebusch}\ \emph {et~al.}(1996)\citenamefont
  {Schlebusch}, \citenamefont {Kessler}, \citenamefont {Cramm},\ and\
  \citenamefont {Eberhardt}}]{Schlebusch1996}%
  \BibitemOpen
  \bibfield  {author} {\bibinfo {author} {\bibfnamefont {C.}~\bibnamefont
  {Schlebusch}}, \bibinfo {author} {\bibfnamefont {B.}~\bibnamefont {Kessler}},
  \bibinfo {author} {\bibfnamefont {S.}~\bibnamefont {Cramm}}, \ and\ \bibinfo
  {author} {\bibfnamefont {W.}~\bibnamefont {Eberhardt}},\ }\href {\doibase
  http://dx.doi.org/10.1016/0379-6779(96)80077-4} {\bibfield  {journal}
  {\bibinfo  {journal} {Synth. Met.}\ }\textbf {\bibinfo {volume} {77}},\
  \bibinfo {pages} {151 } (\bibinfo {year} {1996})}\BibitemShut {NoStop}%
\bibitem [{\citenamefont {Schlebusch}\ \emph {et~al.}(1999)\citenamefont
  {Schlebusch}, \citenamefont {Morenzin}, \citenamefont {Kessler},\ and\
  \citenamefont {Eberhardt}}]{Schlebusch1999}%
  \BibitemOpen
  \bibfield  {author} {\bibinfo {author} {\bibfnamefont {C.}~\bibnamefont
  {Schlebusch}}, \bibinfo {author} {\bibfnamefont {J.}~\bibnamefont
  {Morenzin}}, \bibinfo {author} {\bibfnamefont {B.}~\bibnamefont {Kessler}}, \
  and\ \bibinfo {author} {\bibfnamefont {W.}~\bibnamefont {Eberhardt}},\ }\href
  {\doibase http://dx.doi.org/10.1016/S0008-6223(98)00260-7} {\bibfield
  {journal} {\bibinfo  {journal} {Carbon}\ }\textbf {\bibinfo {volume} {37}},\
  \bibinfo {pages} {717 } (\bibinfo {year} {1999})}\BibitemShut {NoStop}%
\bibitem [{\citenamefont {Morenzin}\ \emph {et~al.}(1999)\citenamefont
  {Morenzin}, \citenamefont {Schlebusch}, \citenamefont {Kessler},\ and\
  \citenamefont {Eberhardt}}]{Morenzin1999}%
  \BibitemOpen
  \bibfield  {author} {\bibinfo {author} {\bibfnamefont {J.}~\bibnamefont
  {Morenzin}}, \bibinfo {author} {\bibfnamefont {C.}~\bibnamefont
  {Schlebusch}}, \bibinfo {author} {\bibfnamefont {B.}~\bibnamefont {Kessler}},
  \ and\ \bibinfo {author} {\bibfnamefont {W.}~\bibnamefont {Eberhardt}},\
  }\href {\doibase 10.1039/A808599D} {\bibfield  {journal} {\bibinfo  {journal}
  {Phys. Chem. Chem. Phys.}\ }\textbf {\bibinfo {volume} {1}},\ \bibinfo
  {pages} {1765} (\bibinfo {year} {1999})}\BibitemShut {NoStop}%
\bibitem [{\citenamefont {Kessler}(1998)}]{Kessler1998}%
  \BibitemOpen
  \bibfield  {author} {\bibinfo {author} {\bibfnamefont {B.}~\bibnamefont
  {Kessler}},\ }\href {\doibase 10.1007/s003390050749} {\bibfield  {journal}
  {\bibinfo  {journal} {Appl. Phys. A}\ }\textbf {\bibinfo {volume} {67}},\
  \bibinfo {pages} {125} (\bibinfo {year} {1998})}\BibitemShut {NoStop}%
\bibitem [{\citenamefont {Wilke}\ \emph {et~al.}(2010)\citenamefont {Wilke},
  \citenamefont {Mizokuro}, \citenamefont {Blum}, \citenamefont {Rabe},\ and\
  \citenamefont {Koch}}]{Wilke2010}%
  \BibitemOpen
  \bibfield  {author} {\bibinfo {author} {\bibfnamefont {A.}~\bibnamefont
  {Wilke}}, \bibinfo {author} {\bibfnamefont {T.}~\bibnamefont {Mizokuro}},
  \bibinfo {author} {\bibfnamefont {R.-P.}\ \bibnamefont {Blum}}, \bibinfo
  {author} {\bibfnamefont {J.~P.}\ \bibnamefont {Rabe}}, \ and\ \bibinfo
  {author} {\bibfnamefont {N.}~\bibnamefont {Koch}},\ }\href {\doibase
  10.1109/JSTQE.2010.2042035} {\bibfield  {journal} {\bibinfo  {journal} {IEEE
  J. Sel. Top. Quant.}\ }\textbf {\bibinfo {volume} {16}},\ \bibinfo {pages}
  {1732} (\bibinfo {year} {2010})}\BibitemShut {NoStop}%
\bibitem [{\citenamefont {Schlebusch}(1998)}]{Schlebusch_thesis}%
  \BibitemOpen
  \bibfield  {author} {\bibinfo {author} {\bibfnamefont {C.}~\bibnamefont
  {Schlebusch}},\ }\href@noop {} {\bibinfo {type} {Thesis}},\ \bibinfo
  {school} {University of Colgne} (\bibinfo {year} {1998})\BibitemShut
  {NoStop}%
\bibitem [{\citenamefont {Gunnarsson}\ \emph {et~al.}(1995)\citenamefont
  {Gunnarsson}, \citenamefont {Handschuh}, \citenamefont {Bechthold},
  \citenamefont {Kessler}, \citenamefont {Gantef\"or},\ and\ \citenamefont
  {Eberhardt}}]{Gunnarsson1995}%
  \BibitemOpen
  \bibfield  {author} {\bibinfo {author} {\bibfnamefont {O.}~\bibnamefont
  {Gunnarsson}}, \bibinfo {author} {\bibfnamefont {H.}~\bibnamefont
  {Handschuh}}, \bibinfo {author} {\bibfnamefont {P.~S.}\ \bibnamefont
  {Bechthold}}, \bibinfo {author} {\bibfnamefont {B.}~\bibnamefont {Kessler}},
  \bibinfo {author} {\bibfnamefont {G.}~\bibnamefont {Gantef\"or}}, \ and\
  \bibinfo {author} {\bibfnamefont {W.}~\bibnamefont {Eberhardt}},\ }\href
  {\doibase 10.1103/PhysRevLett.74.1875} {\bibfield  {journal} {\bibinfo
  {journal} {Phys. Rev. Lett.}\ }\textbf {\bibinfo {volume} {74}},\ \bibinfo
  {pages} {1875} (\bibinfo {year} {1995})}\BibitemShut {NoStop}%
\bibitem [{\citenamefont {Roth}\ \emph {et~al.}(2014)\citenamefont {Roth},
  \citenamefont {Lupulescu}, \citenamefont {Arion}, \citenamefont {Darlatt},
  \citenamefont {Gottwald},\ and\ \citenamefont {Eberhardt}}]{Roth2014}%
  \BibitemOpen
  \bibfield  {author} {\bibinfo {author} {\bibfnamefont {F.}~\bibnamefont
  {Roth}}, \bibinfo {author} {\bibfnamefont {C.}~\bibnamefont {Lupulescu}},
  \bibinfo {author} {\bibfnamefont {T.}~\bibnamefont {Arion}}, \bibinfo
  {author} {\bibfnamefont {E.}~\bibnamefont {Darlatt}}, \bibinfo {author}
  {\bibfnamefont {A.}~\bibnamefont {Gottwald}}, \ and\ \bibinfo {author}
  {\bibfnamefont {W.}~\bibnamefont {Eberhardt}},\ }\href {\doibase
  http://dx.doi.org/10.1063/1.4861886} {\bibfield  {journal} {\bibinfo
  {journal} {J. Appl. Phys.}\ }\textbf {\bibinfo {volume} {115}},\ \bibinfo
  {eid} {033705} (\bibinfo {year} {2014})}\BibitemShut {NoStop}%
\bibitem [{\citenamefont {Opitz}\ \emph {et~al.}(2013)\citenamefont {Opitz},
  \citenamefont {Frisch}, \citenamefont {Schlesinger}, \citenamefont {Wilke},\
  and\ \citenamefont {Koch}}]{Opitz2013}%
  \BibitemOpen
  \bibfield  {author} {\bibinfo {author} {\bibfnamefont {A.}~\bibnamefont
  {Opitz}}, \bibinfo {author} {\bibfnamefont {J.}~\bibnamefont {Frisch}},
  \bibinfo {author} {\bibfnamefont {R.}~\bibnamefont {Schlesinger}}, \bibinfo
  {author} {\bibfnamefont {A.}~\bibnamefont {Wilke}}, \ and\ \bibinfo {author}
  {\bibfnamefont {N.}~\bibnamefont {Koch}},\ }\href
  {http://www.sciencedirect.com/science/article/pii/S0368204812001466}
  {\bibfield  {journal} {\bibinfo  {journal} {J. Electron. Spectrosc. Relat.
  Phenom.}\ }\textbf {\bibinfo {volume} {190}},\ \bibinfo {pages} {12}
  (\bibinfo {year} {2013})},\ \bibinfo {note} {and references
  therein}\BibitemShut {NoStop}%
\bibitem [{\citenamefont {Fahlman}\ \emph {et~al.}(2013)\citenamefont
  {Fahlman}, \citenamefont {Sehati}, \citenamefont {Osikowicz}, \citenamefont
  {Braun}, \citenamefont {de~Jong},\ and\ \citenamefont
  {Brocks}}]{Fahlman2013}%
  \BibitemOpen
  \bibfield  {author} {\bibinfo {author} {\bibfnamefont {M.}~\bibnamefont
  {Fahlman}}, \bibinfo {author} {\bibfnamefont {P.}~\bibnamefont {Sehati}},
  \bibinfo {author} {\bibfnamefont {W.}~\bibnamefont {Osikowicz}}, \bibinfo
  {author} {\bibfnamefont {S.}~\bibnamefont {Braun}}, \bibinfo {author}
  {\bibfnamefont {M.~P.}\ \bibnamefont {de~Jong}}, \ and\ \bibinfo {author}
  {\bibfnamefont {G.}~\bibnamefont {Brocks}},\ }\href
  {http://www.sciencedirect.com/science/article/pii/S0368204813000352}
  {\bibfield  {journal} {\bibinfo  {journal} {J. Electron. Spectrosc. Relat.
  Phenom.}\ }\textbf {\bibinfo {volume} {190}},\ \bibinfo {pages} {33}
  (\bibinfo {year} {2013})},\ \bibinfo {note} {and references
  therein}\BibitemShut {NoStop}%
\bibitem [{\citenamefont {Kushto}\ \emph {et~al.}(2010)\citenamefont {Kushto},
  \citenamefont {M\"akinen},\ and\ \citenamefont {Lane}}]{Kushto2010}%
  \BibitemOpen
  \bibfield  {author} {\bibinfo {author} {\bibfnamefont {G.}~\bibnamefont
  {Kushto}}, \bibinfo {author} {\bibfnamefont {A.}~\bibnamefont {M\"akinen}}, \
  and\ \bibinfo {author} {\bibfnamefont {P.}~\bibnamefont {Lane}},\ }\href
  {\doibase 10.1109/JSTQE.2010.2052354} {\bibfield  {journal} {\bibinfo
  {journal} {IEEE J. Sel. Top. Quant.}\ }\textbf {\bibinfo {volume} {16}},\
  \bibinfo {pages} {1552} (\bibinfo {year} {2010})}\BibitemShut {NoStop}%
\bibitem [{\citenamefont {Akaike}\ \emph {et~al.}(2010)\citenamefont {Akaike},
  \citenamefont {Kanai}, \citenamefont {Ouchi},\ and\ \citenamefont
  {Seki}}]{Akaike2010}%
  \BibitemOpen
  \bibfield  {author} {\bibinfo {author} {\bibfnamefont {K.}~\bibnamefont
  {Akaike}}, \bibinfo {author} {\bibfnamefont {K.}~\bibnamefont {Kanai}},
  \bibinfo {author} {\bibfnamefont {Y.}~\bibnamefont {Ouchi}}, \ and\ \bibinfo
  {author} {\bibfnamefont {K.}~\bibnamefont {Seki}},\ }\href {\doibase
  10.1002/adfm.200901585} {\bibfield  {journal} {\bibinfo  {journal} {Adv.
  Funct. Mater.}\ }\textbf {\bibinfo {volume} {20}},\ \bibinfo {pages} {715}
  (\bibinfo {year} {2010})}\BibitemShut {NoStop}%
\bibitem [{\citenamefont {Koch}(2012)}]{Koch2012}%
  \BibitemOpen
  \bibfield  {author} {\bibinfo {author} {\bibfnamefont {N.}~\bibnamefont
  {Koch}},\ }\href {\doibase 10.1002/pssr.201206208} {\bibfield  {journal}
  {\bibinfo  {journal} {Phys. Stat. Sol. (RRL)}\ }\textbf {\bibinfo {volume}
  {6}},\ \bibinfo {pages} {277} (\bibinfo {year} {2012})}\BibitemShut {NoStop}%
\bibitem [{\citenamefont {Li}\ \emph {et~al.}(2014)\citenamefont {Li},
  \citenamefont {Yu}, \citenamefont {Zhang}, \citenamefont {Yao}, \citenamefont
  {Wu},\ and\ \citenamefont {Hou}}]{Li2014}%
  \BibitemOpen
  \bibfield  {author} {\bibinfo {author} {\bibfnamefont {W.}~\bibnamefont
  {Li}}, \bibinfo {author} {\bibfnamefont {H.}~\bibnamefont {Yu}}, \bibinfo
  {author} {\bibfnamefont {J.}~\bibnamefont {Zhang}}, \bibinfo {author}
  {\bibfnamefont {Y.}~\bibnamefont {Yao}}, \bibinfo {author} {\bibfnamefont
  {C.}~\bibnamefont {Wu}}, \ and\ \bibinfo {author} {\bibfnamefont
  {X.}~\bibnamefont {Hou}},\ }\href {\doibase 10.1021/jp501078f} {\bibfield
  {journal} {\bibinfo  {journal} {J. Phys. Chem. C}\ }\textbf {\bibinfo
  {volume} {118}},\ \bibinfo {pages} {11928} (\bibinfo {year}
  {2014})}\BibitemShut {NoStop}%
\bibitem [{\citenamefont {Castet}\ \emph {et~al.}(2014)\citenamefont {Castet},
  \citenamefont {D{'}Avino}, \citenamefont {Muccioli}, \citenamefont {Cornil},\
  and\ \citenamefont {Beljonne}}]{Castet2014}%
  \BibitemOpen
  \bibfield  {author} {\bibinfo {author} {\bibfnamefont {F.}~\bibnamefont
  {Castet}}, \bibinfo {author} {\bibfnamefont {G.}~\bibnamefont {D{'}Avino}},
  \bibinfo {author} {\bibfnamefont {L.}~\bibnamefont {Muccioli}}, \bibinfo
  {author} {\bibfnamefont {J.}~\bibnamefont {Cornil}}, \ and\ \bibinfo {author}
  {\bibfnamefont {D.}~\bibnamefont {Beljonne}},\ }\href {\doibase
  10.1039/C4CP01872A} {\bibfield  {journal} {\bibinfo  {journal} {Phys. Chem.
  Chem. Phys.}\ }\textbf {\bibinfo {volume} {16}},\ \bibinfo {pages} {20279}
  (\bibinfo {year} {2014})}\BibitemShut {NoStop}%
\bibitem [{\citenamefont {Beltran}\ \emph {et~al.}(2014)\citenamefont
  {Beltran}, \citenamefont {Flores},\ and\ \citenamefont
  {Ortega}}]{Beltran2014}%
  \BibitemOpen
  \bibfield  {author} {\bibinfo {author} {\bibfnamefont {J.}~\bibnamefont
  {Beltran}}, \bibinfo {author} {\bibfnamefont {F.}~\bibnamefont {Flores}}, \
  and\ \bibinfo {author} {\bibfnamefont {J.}~\bibnamefont {Ortega}},\ }\href
  {\doibase 10.1039/C3CP55004D} {\bibfield  {journal} {\bibinfo  {journal}
  {Phys. Chem. Chem. Phys.}\ }\textbf {\bibinfo {volume} {16}},\ \bibinfo
  {pages} {4268} (\bibinfo {year} {2014})}\BibitemShut {NoStop}%
\bibitem [{\citenamefont {Sai}\ \emph {et~al.}(2012)\citenamefont {Sai},
  \citenamefont {Gearba}, \citenamefont {Dolocan}, \citenamefont {Tritsch},
  \citenamefont {Chan}, \citenamefont {Chelikowsky}, \citenamefont {Leung},\
  and\ \citenamefont {Zhu}}]{Sai2012}%
  \BibitemOpen
  \bibfield  {author} {\bibinfo {author} {\bibfnamefont {N.}~\bibnamefont
  {Sai}}, \bibinfo {author} {\bibfnamefont {R.}~\bibnamefont {Gearba}},
  \bibinfo {author} {\bibfnamefont {A.}~\bibnamefont {Dolocan}}, \bibinfo
  {author} {\bibfnamefont {J.~R.}\ \bibnamefont {Tritsch}}, \bibinfo {author}
  {\bibfnamefont {W.-L.}\ \bibnamefont {Chan}}, \bibinfo {author}
  {\bibfnamefont {J.~R.}\ \bibnamefont {Chelikowsky}}, \bibinfo {author}
  {\bibfnamefont {K.}~\bibnamefont {Leung}}, \ and\ \bibinfo {author}
  {\bibfnamefont {X.}~\bibnamefont {Zhu}},\ }\href
  {http://dx.doi.org/10.1021/jz300744r} {\bibfield  {journal} {\bibinfo
  {journal} {J. Phys. Chem. Lett.}\ }\textbf {\bibinfo {volume} {3}},\ \bibinfo
  {pages} {2173} (\bibinfo {year} {2012})}\BibitemShut {NoStop}%
\bibitem [{\citenamefont {Dutton}\ and\ \citenamefont
  {Robey}(2012)}]{Dutton2012}%
  \BibitemOpen
  \bibfield  {author} {\bibinfo {author} {\bibfnamefont {G.~J.}\ \bibnamefont
  {Dutton}}\ and\ \bibinfo {author} {\bibfnamefont {S.~W.}\ \bibnamefont
  {Robey}},\ }\href {\doibase 10.1021/jp305637r} {\bibfield  {journal}
  {\bibinfo  {journal} {J. Phys. Chem. C}\ }\textbf {\bibinfo {volume} {116}},\
  \bibinfo {pages} {19173} (\bibinfo {year} {2012})}\BibitemShut {NoStop}%
\bibitem [{\citenamefont {Jailaubekov}\ \emph {et~al.}(2013)\citenamefont
  {Jailaubekov}, \citenamefont {Willard}, \citenamefont {Tritsch},
  \citenamefont {Chan}, \citenamefont {Sai}, \citenamefont {Gearba},
  \citenamefont {Kaake}, \citenamefont {Williams}, \citenamefont {Leung},
  \citenamefont {Rossky},\ and\ \citenamefont {Zhu}}]{Jailaubekov2013}%
  \BibitemOpen
  \bibfield  {author} {\bibinfo {author} {\bibfnamefont {A.~E.}\ \bibnamefont
  {Jailaubekov}}, \bibinfo {author} {\bibfnamefont {A.~P.}\ \bibnamefont
  {Willard}}, \bibinfo {author} {\bibfnamefont {J.~R.}\ \bibnamefont
  {Tritsch}}, \bibinfo {author} {\bibfnamefont {W.-L.}\ \bibnamefont {Chan}},
  \bibinfo {author} {\bibfnamefont {N.}~\bibnamefont {Sai}}, \bibinfo {author}
  {\bibfnamefont {R.}~\bibnamefont {Gearba}}, \bibinfo {author} {\bibfnamefont
  {L.~G.}\ \bibnamefont {Kaake}}, \bibinfo {author} {\bibfnamefont {K.~J.}\
  \bibnamefont {Williams}}, \bibinfo {author} {\bibfnamefont {K.}~\bibnamefont
  {Leung}}, \bibinfo {author} {\bibfnamefont {P.~J.}\ \bibnamefont {Rossky}}, \
  and\ \bibinfo {author} {\bibfnamefont {X.-Y.}\ \bibnamefont {Zhu}},\ }\href
  {http://dx.doi.org/10.1038/nmat3500} {\bibfield  {journal} {\bibinfo
  {journal} {Nat. Mater}\ }\textbf {\bibinfo {volume} {12}},\ \bibinfo {pages}
  {66} (\bibinfo {year} {2013})}\BibitemShut {NoStop}%
\bibitem [{\citenamefont {Bluhm}\ \emph {et~al.}(2006)\citenamefont {Bluhm},
  \citenamefont {Andersson}, \citenamefont {Araki}, \citenamefont {Benzerara},
  \citenamefont {Brown}, \citenamefont {Dynes}, \citenamefont {Ghosal},
  \citenamefont {Gilles}, \citenamefont {Hansen}, \citenamefont {Hemminger},
  \citenamefont {Hitchcock}, \citenamefont {Ketteler}, \citenamefont
  {Kilcoyne}, \citenamefont {Kneedler}, \citenamefont {Lawrence}, \citenamefont
  {Leppard}, \citenamefont {Majzlam}, \citenamefont {Mun}, \citenamefont
  {Myneni}, \citenamefont {Nilsson}, \citenamefont {Ogasawara}, \citenamefont
  {Ogletree}, \citenamefont {Pecher}, \citenamefont {Salmeron}, \citenamefont
  {Shuh}, \citenamefont {Tonner}, \citenamefont {Tyliszczak}, \citenamefont
  {Warwick},\ and\ \citenamefont {Yoon}}]{Bluhm2006}%
  \BibitemOpen
  \bibfield  {author} {\bibinfo {author} {\bibfnamefont {H.}~\bibnamefont
  {Bluhm}}, \bibinfo {author} {\bibfnamefont {K.}~\bibnamefont {Andersson}},
  \bibinfo {author} {\bibfnamefont {T.}~\bibnamefont {Araki}}, \bibinfo
  {author} {\bibfnamefont {K.}~\bibnamefont {Benzerara}}, \bibinfo {author}
  {\bibfnamefont {G.}~\bibnamefont {Brown}}, \bibinfo {author} {\bibfnamefont
  {J.}~\bibnamefont {Dynes}}, \bibinfo {author} {\bibfnamefont
  {S.}~\bibnamefont {Ghosal}}, \bibinfo {author} {\bibfnamefont
  {M.}~\bibnamefont {Gilles}}, \bibinfo {author} {\bibfnamefont {H.-C.}\
  \bibnamefont {Hansen}}, \bibinfo {author} {\bibfnamefont {J.}~\bibnamefont
  {Hemminger}}, \bibinfo {author} {\bibfnamefont {A.}~\bibnamefont
  {Hitchcock}}, \bibinfo {author} {\bibfnamefont {G.}~\bibnamefont {Ketteler}},
  \bibinfo {author} {\bibfnamefont {A.}~\bibnamefont {Kilcoyne}}, \bibinfo
  {author} {\bibfnamefont {E.}~\bibnamefont {Kneedler}}, \bibinfo {author}
  {\bibfnamefont {J.}~\bibnamefont {Lawrence}}, \bibinfo {author}
  {\bibfnamefont {G.}~\bibnamefont {Leppard}}, \bibinfo {author} {\bibfnamefont
  {J.}~\bibnamefont {Majzlam}}, \bibinfo {author} {\bibfnamefont
  {B.}~\bibnamefont {Mun}}, \bibinfo {author} {\bibfnamefont {S.}~\bibnamefont
  {Myneni}}, \bibinfo {author} {\bibfnamefont {A.}~\bibnamefont {Nilsson}},
  \bibinfo {author} {\bibfnamefont {H.}~\bibnamefont {Ogasawara}}, \bibinfo
  {author} {\bibfnamefont {D.}~\bibnamefont {Ogletree}}, \bibinfo {author}
  {\bibfnamefont {K.}~\bibnamefont {Pecher}}, \bibinfo {author} {\bibfnamefont
  {M.}~\bibnamefont {Salmeron}}, \bibinfo {author} {\bibfnamefont
  {D.}~\bibnamefont {Shuh}}, \bibinfo {author} {\bibfnamefont {B.}~\bibnamefont
  {Tonner}}, \bibinfo {author} {\bibfnamefont {T.}~\bibnamefont {Tyliszczak}},
  \bibinfo {author} {\bibfnamefont {T.}~\bibnamefont {Warwick}}, \ and\
  \bibinfo {author} {\bibfnamefont {T.}~\bibnamefont {Yoon}},\ }\href {\doibase
  http://dx.doi.org/10.1016/j.elspec.2005.07.005} {\bibfield  {journal}
  {\bibinfo  {journal} {J. Electron. Spectrosc. Relat. Phenom.}\ }\textbf
  {\bibinfo {volume} {150}},\ \bibinfo {pages} {86 } (\bibinfo {year}
  {2006})}\BibitemShut {NoStop}%
\bibitem [{\citenamefont {Neppl}\ \emph {et~al.}(2014)\citenamefont {Neppl},
  \citenamefont {Shavorskiy}, \citenamefont {Zegkinoglou}, \citenamefont
  {Fraund}, \citenamefont {Slaughter}, \citenamefont {Troy}, \citenamefont
  {Ziemkiewicz}, \citenamefont {Ahmed}, \citenamefont {Gul}, \citenamefont
  {Rude}, \citenamefont {Zhang}, \citenamefont {Tremsin}, \citenamefont
  {Glans}, \citenamefont {Liu}, \citenamefont {Wu}, \citenamefont {Guo},
  \citenamefont {Salmeron}, \citenamefont {Bluhm},\ and\ \citenamefont
  {Gessner}}]{Neppl2014}%
  \BibitemOpen
  \bibfield  {author} {\bibinfo {author} {\bibfnamefont {S.}~\bibnamefont
  {Neppl}}, \bibinfo {author} {\bibfnamefont {A.}~\bibnamefont {Shavorskiy}},
  \bibinfo {author} {\bibfnamefont {I.}~\bibnamefont {Zegkinoglou}}, \bibinfo
  {author} {\bibfnamefont {M.}~\bibnamefont {Fraund}}, \bibinfo {author}
  {\bibfnamefont {D.~S.}\ \bibnamefont {Slaughter}}, \bibinfo {author}
  {\bibfnamefont {T.}~\bibnamefont {Troy}}, \bibinfo {author} {\bibfnamefont
  {M.~P.}\ \bibnamefont {Ziemkiewicz}}, \bibinfo {author} {\bibfnamefont
  {M.}~\bibnamefont {Ahmed}}, \bibinfo {author} {\bibfnamefont
  {S.}~\bibnamefont {Gul}}, \bibinfo {author} {\bibfnamefont {B.}~\bibnamefont
  {Rude}}, \bibinfo {author} {\bibfnamefont {J.~Z.}\ \bibnamefont {Zhang}},
  \bibinfo {author} {\bibfnamefont {A.~S.}\ \bibnamefont {Tremsin}}, \bibinfo
  {author} {\bibfnamefont {P.-A.}\ \bibnamefont {Glans}}, \bibinfo {author}
  {\bibfnamefont {Y.-S.}\ \bibnamefont {Liu}}, \bibinfo {author} {\bibfnamefont
  {C.~H.}\ \bibnamefont {Wu}}, \bibinfo {author} {\bibfnamefont
  {J.}~\bibnamefont {Guo}}, \bibinfo {author} {\bibfnamefont {M.}~\bibnamefont
  {Salmeron}}, \bibinfo {author} {\bibfnamefont {H.}~\bibnamefont {Bluhm}}, \
  and\ \bibinfo {author} {\bibfnamefont {O.}~\bibnamefont {Gessner}},\ }\href
  {\doibase 10.1039/C4FD00036F} {\bibfield  {journal} {\bibinfo  {journal}
  {Faraday Discuss.}\ }\textbf {\bibinfo {volume} {171}},\ \bibinfo {pages}
  {219} (\bibinfo {year} {2014})}\BibitemShut {NoStop}%
\bibitem [{\citenamefont {Glover}\ \emph {et~al.}(2001)\citenamefont {Glover},
  \citenamefont {Ackermann}, \citenamefont {Belkacem}, \citenamefont
  {Feinberg}, \citenamefont {Heimann}, \citenamefont {Hussain}, \citenamefont
  {Padmore}, \citenamefont {Ray}, \citenamefont {Schoenlein},\ and\
  \citenamefont {Steele}}]{Glover2001}%
  \BibitemOpen
  \bibfield  {author} {\bibinfo {author} {\bibfnamefont {T.}~\bibnamefont
  {Glover}}, \bibinfo {author} {\bibfnamefont {G.}~\bibnamefont {Ackermann}},
  \bibinfo {author} {\bibfnamefont {A.}~\bibnamefont {Belkacem}}, \bibinfo
  {author} {\bibfnamefont {B.}~\bibnamefont {Feinberg}}, \bibinfo {author}
  {\bibfnamefont {P.}~\bibnamefont {Heimann}}, \bibinfo {author} {\bibfnamefont
  {Z.}~\bibnamefont {Hussain}}, \bibinfo {author} {\bibfnamefont
  {H.}~\bibnamefont {Padmore}}, \bibinfo {author} {\bibfnamefont
  {C.}~\bibnamefont {Ray}}, \bibinfo {author} {\bibfnamefont {R.}~\bibnamefont
  {Schoenlein}}, \ and\ \bibinfo {author} {\bibfnamefont {W.}~\bibnamefont
  {Steele}},\ }\href@noop {} {\bibfield  {journal} {\bibinfo  {journal} {Nucl.
  Instrum. Methods Phys. Res. Sec. A}\ }\textbf {\bibinfo {volume} {467}},\
  \bibinfo {pages} {1438} (\bibinfo {year} {2001})}\BibitemShut {NoStop}%
\bibitem [{\citenamefont {Laurs}\ and\ \citenamefont
  {Heiland}(1987)}]{Laurs1987}%
  \BibitemOpen
  \bibfield  {author} {\bibinfo {author} {\bibfnamefont {H.}~\bibnamefont
  {Laurs}}\ and\ \bibinfo {author} {\bibfnamefont {G.}~\bibnamefont
  {Heiland}},\ }\href {\doibase http://dx.doi.org/10.1016/0040-6090(87)90288-4}
  {\bibfield  {journal} {\bibinfo  {journal} {Thin Solid Films}\ }\textbf
  {\bibinfo {volume} {149}},\ \bibinfo {pages} {129 } (\bibinfo {year}
  {1987})}\BibitemShut {NoStop}%
\bibitem [{\citenamefont {Long}\ \emph {et~al.}(1990)\citenamefont {Long},
  \citenamefont {Sadeghi}, \citenamefont {Rife},\ and\ \citenamefont
  {Kabler}}]{Long1990}%
  \BibitemOpen
  \bibfield  {author} {\bibinfo {author} {\bibfnamefont {J.~P.}\ \bibnamefont
  {Long}}, \bibinfo {author} {\bibfnamefont {H.~R.}\ \bibnamefont {Sadeghi}},
  \bibinfo {author} {\bibfnamefont {J.~C.}\ \bibnamefont {Rife}}, \ and\
  \bibinfo {author} {\bibfnamefont {M.~N.}\ \bibnamefont {Kabler}},\ }\href
  {\doibase 10.1103/PhysRevLett.64.1158} {\bibfield  {journal} {\bibinfo
  {journal} {Phys. Rev. Lett.}\ }\textbf {\bibinfo {volume} {64}},\ \bibinfo
  {pages} {1158} (\bibinfo {year} {1990})}\BibitemShut {NoStop}%
\bibitem [{\citenamefont {Poirier}\ \emph {et~al.}(1995)\citenamefont
  {Poirier}, \citenamefont {Owens},\ and\ \citenamefont
  {Weaver}}]{Poirier1995}%
  \BibitemOpen
  \bibfield  {author} {\bibinfo {author} {\bibfnamefont {D.~M.}\ \bibnamefont
  {Poirier}}, \bibinfo {author} {\bibfnamefont {D.~W.}\ \bibnamefont {Owens}},
  \ and\ \bibinfo {author} {\bibfnamefont {J.~H.}\ \bibnamefont {Weaver}},\
  }\href {\doibase 10.1103/PhysRevB.51.1830} {\bibfield  {journal} {\bibinfo
  {journal} {Phys. Rev. B}\ }\textbf {\bibinfo {volume} {51}},\ \bibinfo
  {pages} {1830} (\bibinfo {year} {1995})}\BibitemShut {NoStop}%
\bibitem [{\citenamefont {Haddon}(1992)}]{Haddon1992}%
  \BibitemOpen
  \bibfield  {author} {\bibinfo {author} {\bibfnamefont {R.~C.}\ \bibnamefont
  {Haddon}},\ }\href {\doibase 10.1021/ar00015a005} {\bibfield  {journal}
  {\bibinfo  {journal} {Acc. Chem. Res.}\ }\textbf {\bibinfo {volume} {25}},\
  \bibinfo {pages} {127} (\bibinfo {year} {1992})}\BibitemShut {NoStop}%
\bibitem [{\citenamefont {Davidson}(1982)}]{Davidson1982}%
  \BibitemOpen
  \bibfield  {author} {\bibinfo {author} {\bibfnamefont {A.~T.}\ \bibnamefont
  {Davidson}},\ }\href {\doibase http://dx.doi.org/10.1063/1.443636} {\bibfield
   {journal} {\bibinfo  {journal} {J. Chem. Phys.}\ }\textbf {\bibinfo {volume}
  {77}},\ \bibinfo {pages} {168} (\bibinfo {year} {1982})}\BibitemShut
  {NoStop}%
\bibitem [{\citenamefont {Farag}(2007)}]{Farag2007}%
  \BibitemOpen
  \bibfield  {author} {\bibinfo {author} {\bibfnamefont {A.~A.~M.}\
  \bibnamefont {Farag}},\ }\href {\doibase
  http://dx.doi.org/10.1016/j.optlastec.2006.03.011} {\bibfield  {journal}
  {\bibinfo  {journal} {Opt. Las. Tech.}\ }\textbf {\bibinfo {volume} {39}},\
  \bibinfo {pages} {728} (\bibinfo {year} {2007})}\BibitemShut {NoStop}%
\bibitem [{\citenamefont {Kamiya~Okudaira}\ \emph {et~al.}(1999)\citenamefont
  {Kamiya~Okudaira}, \citenamefont {Hasegawa}, \citenamefont {Ishii},
  \citenamefont {Seki}, \citenamefont {Harada},\ and\ \citenamefont
  {Ueno}}]{Ueno1999}%
  \BibitemOpen
  \bibfield  {author} {\bibinfo {author} {\bibfnamefont {K.}~\bibnamefont
  {Kamiya~Okudaira}}, \bibinfo {author} {\bibfnamefont {S.}~\bibnamefont
  {Hasegawa}}, \bibinfo {author} {\bibfnamefont {H.}~\bibnamefont {Ishii}},
  \bibinfo {author} {\bibfnamefont {K.}~\bibnamefont {Seki}}, \bibinfo {author}
  {\bibfnamefont {Y.}~\bibnamefont {Harada}}, \ and\ \bibinfo {author}
  {\bibfnamefont {N.}~\bibnamefont {Ueno}},\ }\href {\doibase
  http://dx.doi.org/10.1063/1.370149} {\bibfield  {journal} {\bibinfo
  {journal} {J. Appl. Phys.}\ }\textbf {\bibinfo {volume} {85}},\ \bibinfo
  {pages} {6453} (\bibinfo {year} {1999})}\BibitemShut {NoStop}%
\bibitem [{\citenamefont {Schwieger}\ \emph {et~al.}(2002)\citenamefont
  {Schwieger}, \citenamefont {Peisert}, \citenamefont {Golden}, \citenamefont
  {Knupfer},\ and\ \citenamefont {Fink}}]{Schwieger2002}%
  \BibitemOpen
  \bibfield  {author} {\bibinfo {author} {\bibfnamefont {T.}~\bibnamefont
  {Schwieger}}, \bibinfo {author} {\bibfnamefont {H.}~\bibnamefont {Peisert}},
  \bibinfo {author} {\bibfnamefont {M.~S.}\ \bibnamefont {Golden}}, \bibinfo
  {author} {\bibfnamefont {M.}~\bibnamefont {Knupfer}}, \ and\ \bibinfo
  {author} {\bibfnamefont {J.}~\bibnamefont {Fink}},\ }\href {\doibase
  10.1103/PhysRevB.66.155207} {\bibfield  {journal} {\bibinfo  {journal} {Phys.
  Rev. B}\ }\textbf {\bibinfo {volume} {66}},\ \bibinfo {pages} {155207}
  (\bibinfo {year} {2002})}\BibitemShut {NoStop}%
\end{thebibliography}
\end{document}